\documentclass[twocolumn,superscriptaddress,prl,longbibliography]{revtex4-1}
\usepackage{graphicx}
\usepackage{amssymb}
\usepackage{amsmath}
\usepackage{epsfig}
\usepackage{color}
\usepackage{mathtools}
\usepackage[colorlinks,linkcolor=blue,anchorcolor=blue,citecolor=blue,urlcolor=blue]{hyperref}

\begin{document}
\title{Focus the electromagnetic field to $10^{-6} \lambda$ for ultra-high enhancement of field-matter interaction}

\author{Xiang-Dong Chen}
\author{En-Hui Wang}
\author{Long-Kun Shan}
\author{Ce Feng}
\author{Yu Zheng}
\author{Yang Dong}
\author{Guang-Can Guo}
\author{Fang-Wen Sun}
\email{fwsun@ustc.edu.cn}

\affiliation{CAS Key Laboratory of Quantum Information, School of Physical Sciences, University of Science and Technology of China, Hefei, 230026, People's Republic of China}
\affiliation{CAS Center For Excellence in Quantum Information and Quantum Physics, University of Science and Technology of China, Hefei, 230026, People's Republic of China}
\date{\today}

\begin{abstract}
Focusing electromagnetic field to enhance the interaction with matter has been promoting researches and applications of nano electronics and photonics.
Usually, the evanescent-wave coupling is adopted in various nano structures and materials to confine the electromagnetic field into a subwavelength space.
Here, based on the direct coupling with confined electron oscillations in a nanowire, we demonstrate an extreme localization of microwave field down to 10$^{-6}\lambda$.
A hybrid nanowire-bowtie antenna is further designed to focus the free-space microwave to this deep-subwavelength space.
Detected by the nitrogen vacancy center in diamond, the field intensity and microwave-spin interaction strength are enhanced by 2.0$\times$10$^{8}$ and 1.4$\times$10$^{4}$ times, respectively.
Such an extreme concentration of microwave field will further promote integrated quantum information processing, sensing and microwave photonics in a nanoscale system.
\end{abstract}

\maketitle

Electromagnetic field can usually be focused at the scale of its wavelength. However, in pursuit of a strong interaction with matter, the manipulation of the electromagnetic field in a subwavelength space is one of the most important tasks in nanoscience researches and applications, ranging from integrated optics to biological
sensing\cite{marpaung2019nature-IMWP,GNP-sensing,cheben2018subwavelength,kauranen2012nprev}.
Nanostructures of dielectric\cite{weiss-diedeep-2018sciadv,Englund-nanocavity-prl2017,Igor-diele-2020nl}, metallic\cite{lee-plaante-2015nl,Baumberg-plasnano2016nat,xuhx2018probing} and two-dimensional materials\cite{Koppens-2dplas-2018science,xia2014two} have been developed to tightly confine the electromagnetic field mainly based on the evanescent-wave coupling. For example, the plasmonic nanostructure has been used for the light field confinement at a scale smaller than $10^{-2}\lambda$\cite{lee-plaante-2015nl,Baumberg-plasnano2016nat,xuhx2018probing}. These confinements can dramatically reduce the mode volume to highly increase the density of states and enhance the light-matter interaction at the nanoscale, which has harnessed the researches of
single molecule spectroscopy\cite{hou-2013nat}, nano laser\cite{azzam2020ten}, nonlinear optics\cite{Smirnova2016nonlinear}, and solar energy\cite{atwater2010plasmonics}.

Especially, the interaction between microwave field and matter at the nanoscale strongly drives the development of quantum information processing, sensing and microwave photonics.
The deep-subwavelength confinement of the microwave field will benefit the individual manipulation of multi-qubit\cite{Wineland-prl2013-MWion,lucas-2016-nearMW}. Meanwhile, the enhancement of local microwave field is of central importance to microwave-to-optics conversion\cite{Leuthold-2018MWP-np,loncar-2018nature-MWP}, fast spin qubit manipulation\cite{aws-sci20101} and hybrid quantum system coupling\cite{clerk2020hybrid,supcon-prl20101}.
It indicates that, the efficient localization and detection of microwave field at the nanoscale is highly required for develop a practical quantum information device. Further more, the wireless qubit manipulation with a compact and scalable system will decrease the power consumption and reduce the heat load in a cryostat\cite{blais-2020nphys-quantum,Reilly-2021ne-cryogenic}.

Here, we study the field confinement based on the direct coupling between electromagnetic field and confined electrons in a low dimensional nano material.
An extreme-confinement of a microwave field with an ultra-strong intensity is realized by utilizing the near field radiation of the electron oscillation in an Ag nanowire.
Using the NV center in diamond as a noninvasive probe, we show that the microwave field can be localized down to 291 nm, corresponding to a scale of $10^{-6} \lambda$. For far-field spin manipulation, we design a hybrid nanowire-bowtie structure to focus the microwave field directly from the free space to a deep-subwavelength volume. As a result, the microwave-spin interaction strength is highly enhanced by observing a 1.4$\times$10$^{4}$ times enhancement of the Rabi oscillation frequency, corresponding to increasing the field intensity by 2.0$\times$10$^{8}$ times. Further considering the light guiding effect of the Ag nanowire, this antenna can be used for delivering and concentrating both light and microwave field. Subsequently, a wireless platform can be developed for the integrated quantum information processing and quantum sensing.

\begin{figure*}
  \centering
  \includegraphics[width=0.9\textwidth]{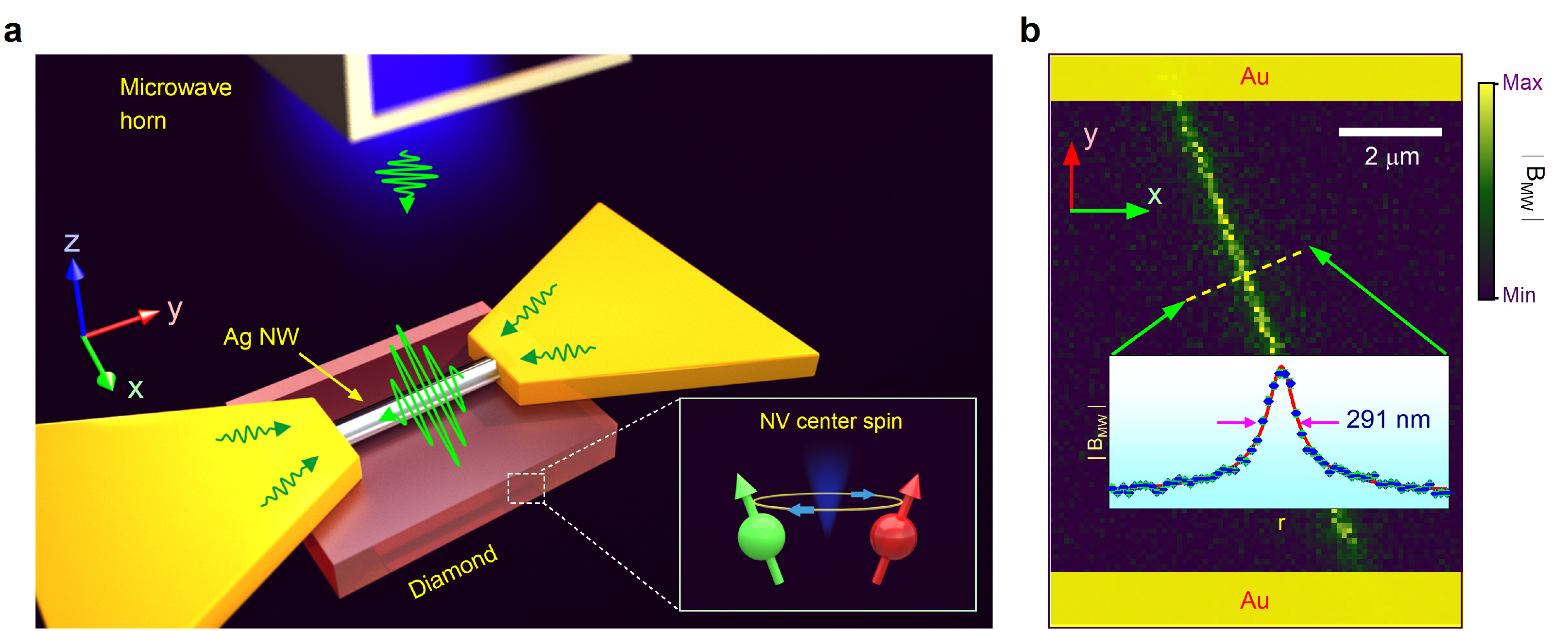}
  \caption{\textbf{The principle of localizing and detecting the microwave field.} (a) Sketch of the nanowire-bowtie antenna. A single crystal diamond plate is placed under the nanowire. The spin state transition of NV center in diamond is pumped by the localized microwave. (b) The image of the microwave distribution is obtained by recording the spin state transition of NV center. The insert is the integrated cross-section profile. The solid line is the fit with equation (\ref{eqbamp}). The power of the microwave that is radiated by the horn antenna is 14 $\mu$W.}\label{figschem}
\end{figure*}

\section{Results}
\subsection{The design of experiments}
As shown in Fig. \ref{figschem}(a), the nanowire-bowtie hybrid antenna consists of an Ag nanowire with a diameter of 120 nm and a metallic bowtie structure. The gap between the two arms of the bowtie structure is $W_{gap} =$ 8 $\mu$m. The length of the bowtie structure is 6.5 cm, while the widths at the end and at the gap are 1 cm and 160 $\mu$m, respectively (details in Supplementary information). A double-ridged horn antenna radiates the microwave signal into the
free space. The nanowire-bowtie structure then receives the far-field microwave (with a distance of approximate 20 cm).

The NV center in a single crystal diamond plate is generated
by nitrogen ion implanting. The depth is approximate 20 nm. The ground
state of NV center is a spin triplet. The transition between the $m_{s}=0$ and $m_{s}=\pm 1$ can be pumped by a resonant microwave. It subsequently changes the fluorescence intensity of NV center. To detect the microwave, we record the optically detected magnetic resonance (ODMR) of NV center under a continuous-wave microwave pumping. The contrast of the ODMR signal increases with the amplitude of the microwave field (Supplementary information).
To non-invasively map the localized microwave with a high spatial resolution, the charge state depletion (CSD) nanoscopy\cite{chen201501} is applied for the diffraction-unlimited ODMR measurement. It is based on the charge state manipulation and detection of NV center. The resolution of CSD nanoscopy is approximate 100 nm here, in comparison with the 500 nm resolution of the confocal microscopy (Supplementary information).

\subsection{Microwave localization and detection}
In Fig. \ref{figschem} (b), we show the magnetic component of the microwave field near the Ag nanowire. Here, without an external magnetic field, the resonant microwave frequency for NV center spin transition is 2.87 GHz. The result shows that the near-field microwave is confined near the Ag nanowire. The width of the cross-section profile is 291$\pm10$ nm, corresponding to 2.8$\times 10^{-6} \lambda$, where $\lambda$ is the microwave wavelength in vacuum.
The distribution of the magnetic component can be well fitted by a reciprocal function:
\begin{equation}\label{eqbamp}
  |B_{MW}|\propto \frac{1}{\sqrt{r^2+r_{0}^2}},
\end{equation}
where $r$ is the distance from the nanowire in the xy plane and $r_0=$84$\pm$3 nm, which is determined by the radius of the nanowire and the depth of NV center. In contrast, the evanescent-field coupling shows an exponential decay as a function of distance\cite{kauranen2012nprev}.

\begin{figure*}
  \centering
  \includegraphics[width=0.8\textwidth]{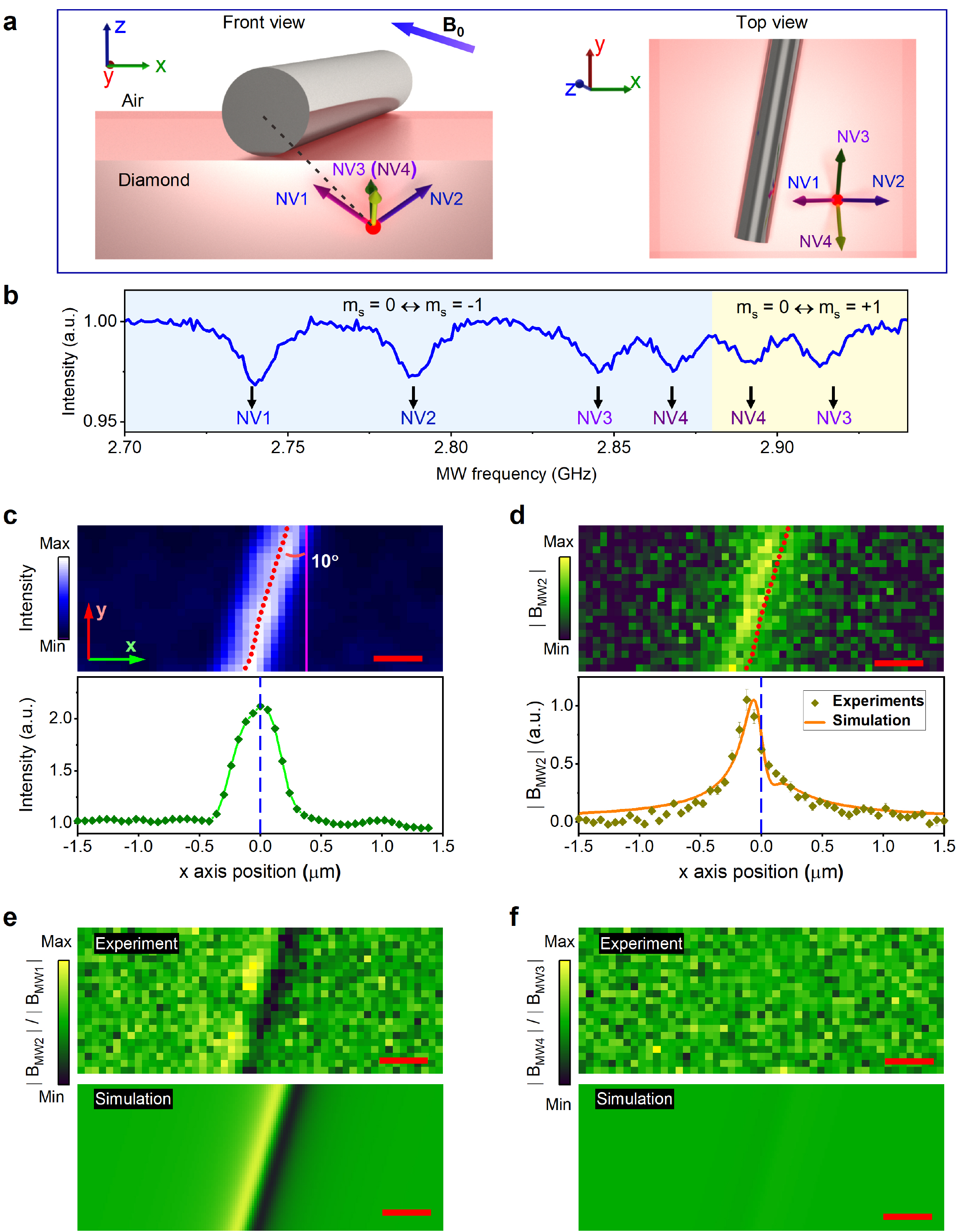}
  \caption{\textbf{Detect the microwave field vector at the nanoscale.} (a) The illustration of the direction of nanowire and NV center axes. (b) Frequency-scanning ODMR results of NV centers with four symmetry axes. (c) The fluorescence intensity of NV centers is enhanced due to the interaction with the Ag nanowire. The cross-section profile in the bottom is used to locate the relative position of nanowire (red points in the image). (d) The image of $|B_{MW2}|$. The cross-section profile is the integrated signal of the whole image. The solid line is the simulation with a straight line current by setting the x axis position of the nanowire to 0. (e)(f) The microwave vector is revealed by comparing different projections. The scale bars in all the images are 400 nm in length. And the power of the microwave source is set to 174 $\mu$W.}\label{figcsd}
\end{figure*}

\begin{figure*}
  \centering
  \includegraphics[width=\textwidth]{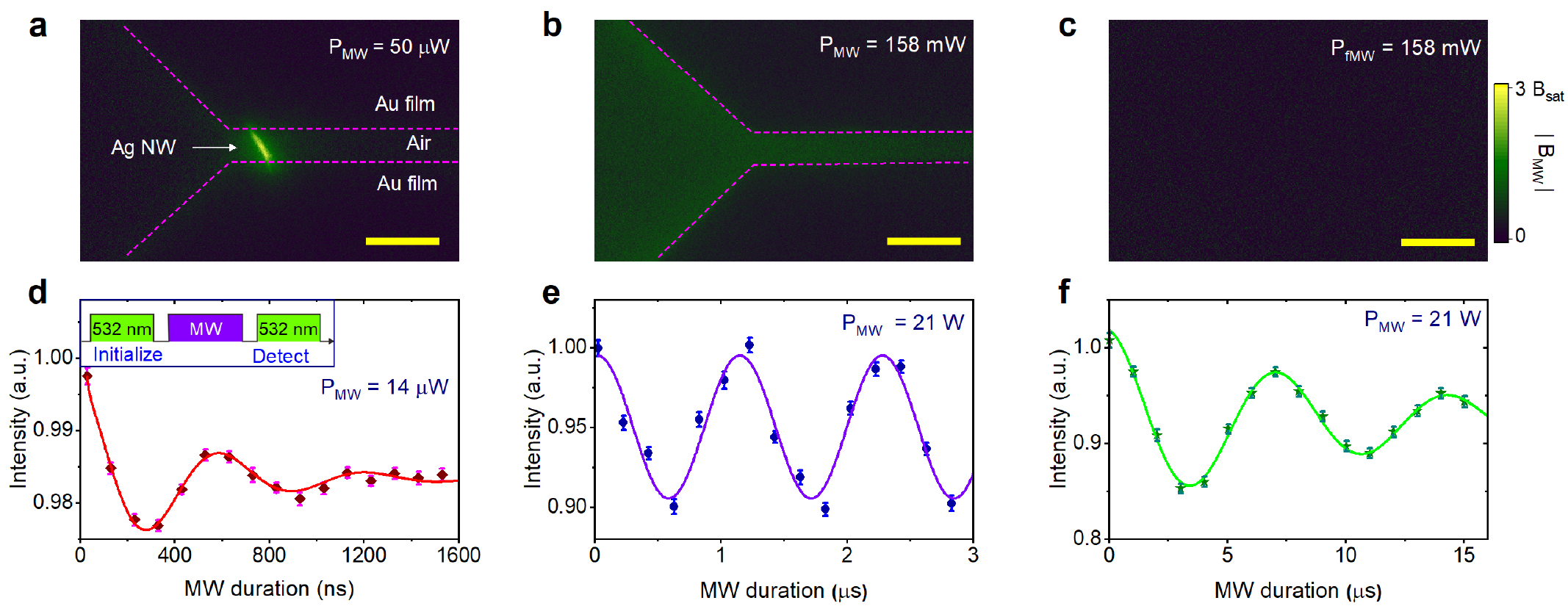}
  \caption{\textbf{Spin manipulation with the localized microwave field.} The wide field imaging illustrates the microwave field enhancement with different structures: (a) nanowire-bowtie antenna; (b) bowtie
  antenna without nanowire; (c) no structure on diamond surface. The scale bars are 20 $\mu$m in length. $B_{sat}$ is the saturation microwave amplitude defined in equation (S2) in Supplementary information. (d)-(f) are Rabi oscillations of NV center with different structures in (a)-(c), respectively. The inset in (d) shows pulse sequences for the Rabi oscillation measurement. The Rabi oscillation in (d) is measured with NV center ensemble under the nanowire. And Rabi oscillations in (e) and (f) are measured with a single NV center. $P_{MW}$ is the power of the microwave source.}\label{figmwimage}
\end{figure*}

To reveal the mechanism of the microwave confinement, we separate the ODMR signal from the four categories of NV centers with different symmetry axes and obtain the vector information of the localized microwave field, as shown in Fig. \ref{figcsd}(a). Here, the x, y, and z axes are defined as the edges of the diamond plate. The symmetry axes of the four categories of NV centers are shown as NV1 (-$\sqrt{2}$, 0, 1), NV2 ($\sqrt{2}$, 0, 1), NV3 (0, $\sqrt{2}$, 1), NV4 (0, -$\sqrt{2}$, 1). An external static magnetic field $\mathbf{B_{0}}$ is applied to split the four categories (NVi), which is shown in Fig. \ref{figcsd}(b). We assume that the localized microwave field distribution with different frequencies is the same. By recording the ODMR signal of NVi centers at different positions, we obtain the distribution of the magnetic component $B_{MWi}$, which is perpendicular to the NVi centers' symmetry axis.

To precisely map the microwave vector distribution, the position of the nanowire is firstly located according to the fluorescence enhancement of NV center, as shown in Fig. \ref{figcsd}(c). We find that the position of the magnetic component's maximum does not always match the position of the nanowire. As shown in Fig. \ref{figcsd}(d), the maximum of $B_{MW2}$ is approximate 100 nm away from the nanowire.
In Figs. \ref{figcsd}(e) and \ref{figcsd}(f), by simultaneously detecting the signals from different NV categories, we further highlight the difference of the distributions with $|B_{MW2}|/|B_{MW1}|$ and $|B_{MW4}|/|B_{MW3}|$, respectively. With the structure in Fig. \ref{figcsd}, the amplitude of $B_{MW3}$  and $B_{MW4}$ are almost same, while the amplitude of $B_{MW1}$  is mirroring  $B_{MW2}$ with respect to the yz plane. The results show that the vector of the magnetic component is in the xz plane. Comparing with the simulation (Supplementary information), it suggests that the localized microwave field is mainly determined by a straight line current that is transmitting through the Ag nanowire. Such a field confinement is not from the evanescent-wave coupling. The small mismatch between the simulation with a straight line current and the experiments might be caused by the error of nanowire's location, the distribution of NV center in z axis, and the accuracy of CSD microscopy resolution's estimation.

\subsection{Far-field spin manipulation}

The extreme confinement leads to a significant enhancement of the localized microwave field's intensity. It subsequently enhances the interaction with a spin qubit. In Fig. \ref{figmwimage}, we compares the localized
microwave field with three different structures: the nanowire-bowtie hybrid antenna, the bowtie antenna without Ag nanowire, and no antenna. The results show that, with a nanowire-bowtie antenna, the microwave is significantly enhanced near the Ag nanowire, while a bowtie
antenna without Ag nanowire slightly increases the localized microwave field in the gap between the two arms.

 \begin{figure*}
  \centering
  \includegraphics[width=0.65\textwidth]{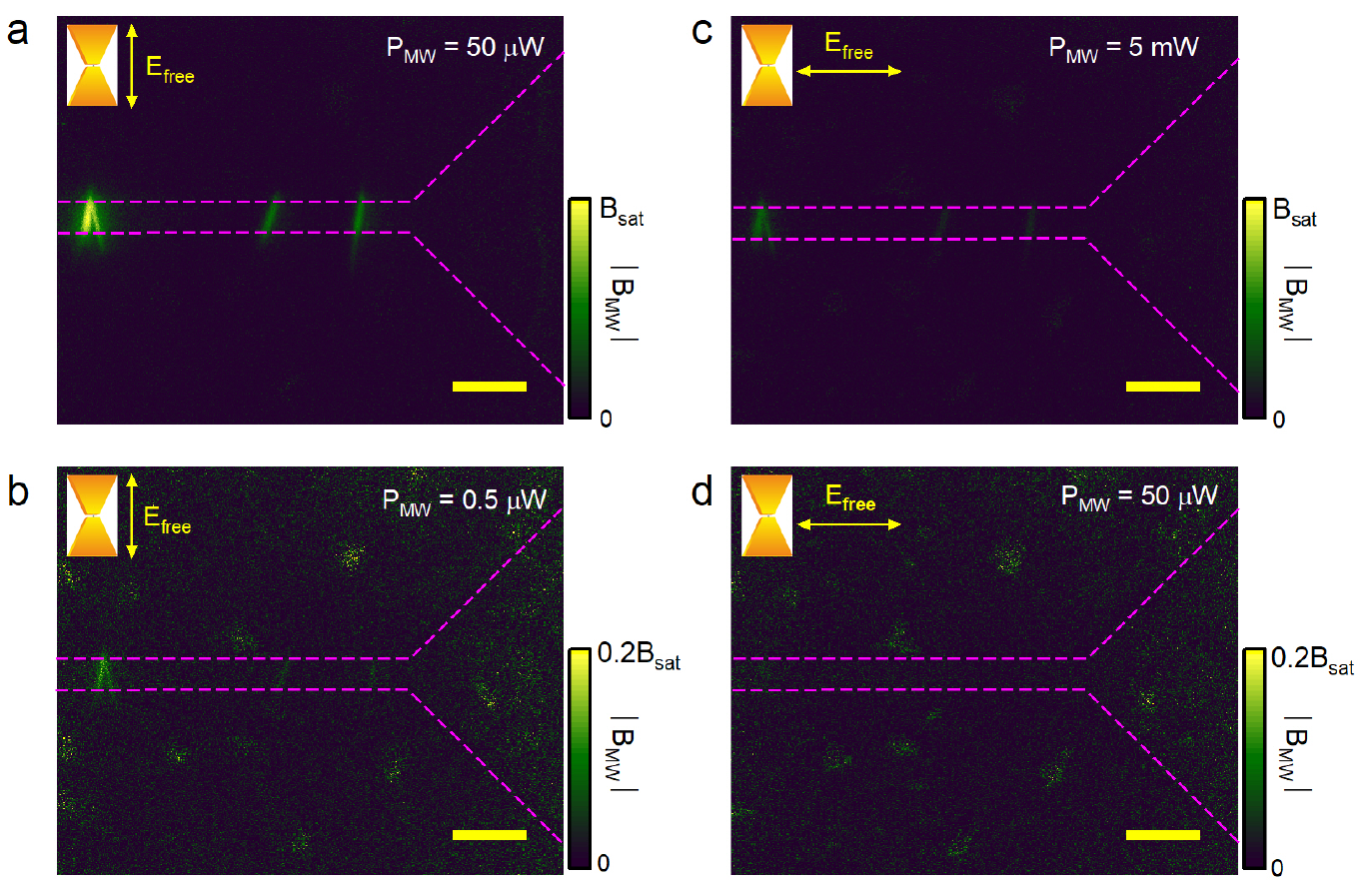}
  \caption{\textbf{The localized microwave field distribution is changed with the polarization of the free-space microwave.} (a)(b) The electrical polarization of the free-space microwave is parallel to the direction of nanowire-bowtie structure. (c)(d) The electrical polarization is perpendicular to the nanowire-bowtie structure. The scale bars are 20 $\mu$m in length. $E_{free}$ denotes the electrical component of the free-space microwave. $P_{MW}$ and $B_{sat}$ are defined as same as that in Fig. \ref{figmwimage}.}\label{figpolar}
\end{figure*}

The extreme microwave field enhancement can be used for fast and high spatial resolution spin qubit manipulation. Here, we use it to pump the Rabi oscillation of NV center. As shown in Fig. \ref{figmwimage}(d), with a nanowire-bowtie antenna,
the Rabi frequency of the NV center under the Ag nanowire is approximate 1.6 $\mu$s$^{-1}$ with a 14 $\mu$W microwave excitation. Without Ag nanowire, the Rabi oscillation frequency of the NV center in the gap of the bowtie antenna is approximate 0.89 $\mu$s$^{-1}$ with a 21 W microwave excitation. In contrast, without any nanostructure, the Rabi oscillation frequency is only 0.14 $\mu$s$^{-1}$ under a 21 W microwave excitation. The results indicate that, by utilizing the nanowire-bowtie antenna for spin manipulation, the Rabi frequency can be improved by at least 1.4$\times 10^{4}$ times, corresponding to increasing the local microwave intensity by 2.0$\times 10^{8}$ times. The observation of Rabi oscillation also indicates that the coherence of both the spin and the localized microwave is preserved, which is crucial for quantum applications.

The individual addressing of multi qubits from far field can be further explored by utilizing the polarization dependence of the microwave localization.
Here, we rotate the horn antenna to change the polarization of the free-space microwave.
The results in Fig. \ref{figpolar} show that, with an electrical polarization parallel to the bowtie-nanowire antenna, a stronger localized microwave field is observed near the nanowire-bowtie antenna. The polarization isolation of the localized microwave intensity with the nanowire-bowtie antenna is higher than 20 dB, but lower than 40 dB. Therefore, encoding the microwave for different spin manipulation into the polarization, the nanowire-bowtie antennas with different directions can be used to selectively manipulate the qubits at different positions from the far-field.

\section{discussion}
Integrating and miniaturizing the electrical and optical device is essential for the practical quantum processing and sensing applications\cite{englund-2019ne,jph-2020nat-integrated}.
Various electrical conductive and ferromagnetic structures have been studied to transmit the microwave signal for spin manipulation at the nanoscale\cite{Staacke2019npj,Bertelli-2020sciadv-FMR}. The enhancement of the nanoscale microwave field has also been demonstrated with an in-plane slotted patch antenna\cite{Leuthold-2018MWP-np}. However, the Johnson noise of a large metal film will cause the spin relaxation \cite{lukin-sci-noise2015,Jayich-johnson-2018nc}, while the Johnson noise from an Ag nanowire is negligible (Supplementary information).
Our method provides a solution to efficient spin manipulation from the far-field. It simplifies the quantum processing device, avoids the thermal leakage in a cryostat.
In addition, the light guiding effect of an Ag nanowire can be utilized to optically pump and collect the fluorescence of individual qubit\cite{lukin-2011prl,andersen-NWdimer-NL2013,xuhongxing-acschemrev2018}. The simultaneous integration of electrical and optical components certainly will help to develop a compact and integrated quantum processing device.

Focusing and detecting the nanoscale electromagnetic field can be further used to enhance the sensitivity of the spin-based metrology. Recently, the magnetic concentrator of ferrite material has been applied for the detection of static magnetic field with NV center\cite{Acosta2019bowtie,wrach-2021-preprint}. The key is how small the electromagnetic field can be focused and detected. Utilizing the nanowire-bowtie antenna, we can improve the sensitivity by 1.4$\times 10^{4}$ times. Combining with the coherent spin manipulation\cite{barry-2019-RMPrevie}, it will help to realize the ultra-weak microwave signal sensing, such as for a quantum radar.

In conclusion, we have demonstrated the extreme concentration of a microwave field by utilizing the confined electron's oscillation in a low dimensional material. The results can be used for integrated quantum information processing and high sensitivity quantum sensing.

\noindent

\section*{Methods}

\noindent
\textbf{Sample preparation.} The electrical grade diamond plates with $\left\{ 100\right\}$ surface and $<$110$>$ edges are purchased from Element 6. The size of the plate is $2\times2\times0.5$ mm$^{3}$. The NV center ensemble is produced through nitrogen ion implanting with an energy of 15 keV and a dosage of 10$^{13}$/cm$^{2}$. The diamond is annealed at 850$^{\circ}$C for 2 hours to improve the production efficiency of NV centers. The density of NV centers is estimated to be approximate 5000/$\mu$m$^2$.
After the NV center is produced, Ag nanowires are dropped on the surface of diamond with a spin processor. Then, a small metallic bowtie structure of chromium/gold (5/200 nm thickness) film is produced on the diamond surface through lift-off. Finally, a large in-plane bowtie antenna is made with a copper foil tap. The Au film on the diamond plate is ohm connected to the copper tape with silver glue.

\begin{figure*}
  \centering
  \includegraphics[width=0.9\textwidth]{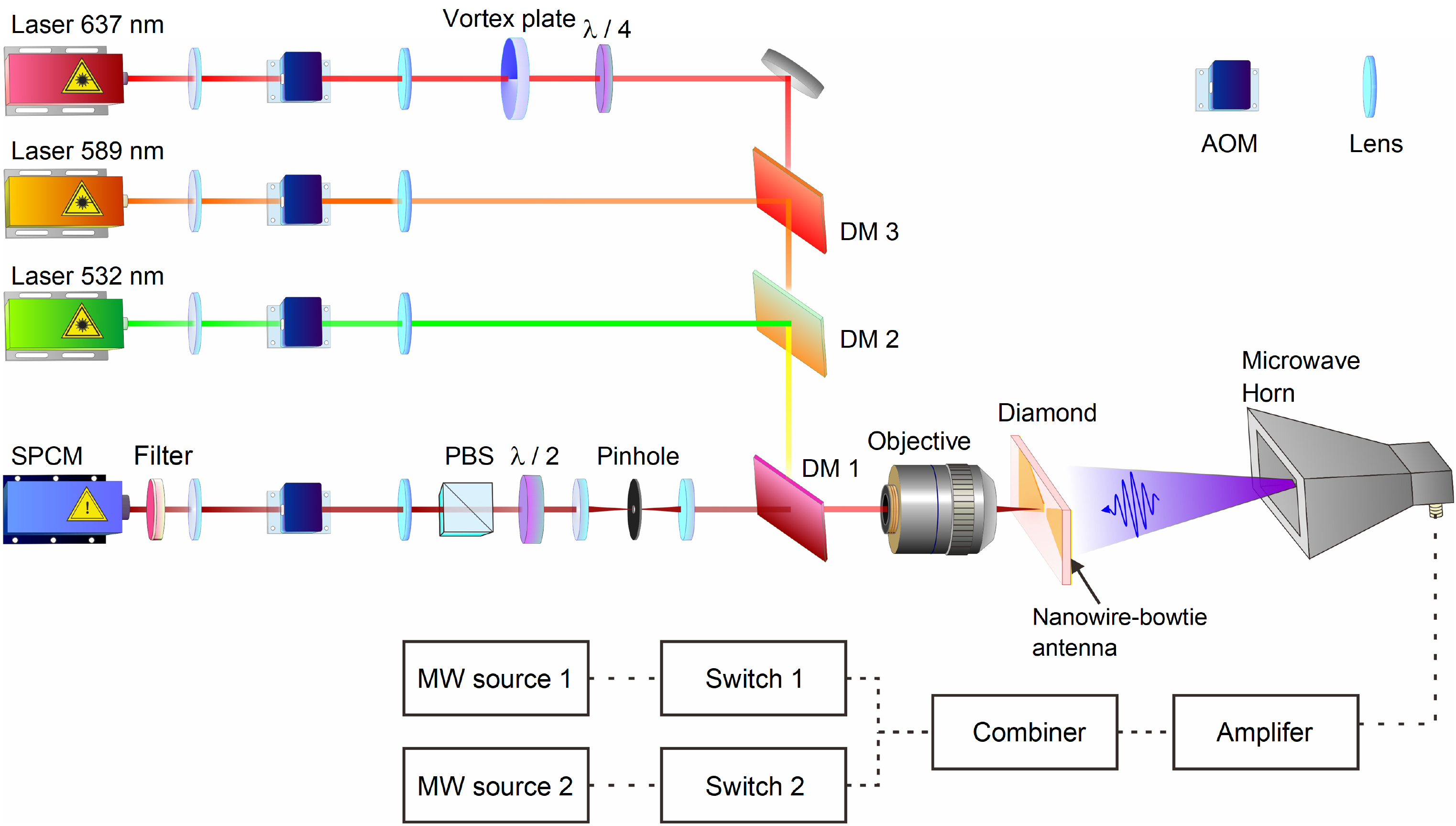}
  \caption{\textbf{The schematic diagram of experimental setup for the CSD nanoscopy.} DM1-3, long-pass dichroic mirror with edge wavelengths of 658.8, 536.8 and 605 nm, respectively; AOM, acousto-optic modulator; SPCM, single-photon-counting modulator; PBS, polarizing beam splitter.}\label{figset}
\end{figure*}

\noindent
\textbf{Experimental setup.} The CSD nanoscopy setup for ODMR measurement is based on a home-built confocal microscope, as shown in Fig. \ref{figset}. Diamond plate is mounted on a piezo-stage (P-733.3DD, PI). CW lasers with wavelengths of 532 (Coherent), 589 (MGL-III-589nm, New Industries Optoelectronics) and 637 nm (MRL-III-637nm, New Industries Optoelectronics) are modulated by acousto-optic modulators (AOMs, MT200-0.5-VIS, AA). A vortex phase plate (VPP-1a, RPC photonics) is used to produce a doughnut-shaped 637 nm laser beam. The lasers pump the NV center in diamond from the backside through an objective (Leica) with 0.7 NA. The collected fluorescence is time gated by another AOM. Then, it is detected by a single-photon-counting-module (SPCM-AQRH-15-FC, Excelitas) after passing through a long-pass filter (edge wavelength 668.9 nm, Semrock). In the wide field microscope, a 532 nm CW laser (MLL-III-532nm, New Industries Optoelectronics) is used to pump the NV center. The fluorescence is detected by a CCD camera (iXon897, Andor).

Two microwave generators (SMB100A and SMA 100A, Rhode\&Schwartz) are used to produce microwave signal with different frequencies. The microwave pulse is controlled by microwave switches (ZASWA-2-50DR, MiniCircuits). Then, the two channels are combined by a combiner (ZFRSC-42-S, MiniCircuits) and amplified by a microwave amplifier (60S1G4A, Amplifier Research). The microwave is radiated into free space by a horn antenna (LB-2080-SF, Chengdu AINFO Inc.).

~\\
\noindent
\textbf{Acknowledgment}

\noindent
This work was supported by National Key Research and
Development Program of China (No. 2017YFA0304504); Anhui Initiative in Quantum Information Technologies (AHY130100); National Natural
Science Foundation of China (Nos. 91536219, 91850102).

\end{document}



\begin{center}

\large
\vspace{15 mm}
\textbf{{Supplementary Information for: \\ \vspace{5 mm} ``Focus the electromagnetic field to $10^{-6} \lambda$ for ultra-high enhancement of field-matter interaction''}}

\normalsize
\vspace{15 mm}
Xiang-Dong Chen$^{1,2}$, En-Hui Wang$^{1,2}$, Long-Kun Shan$^{1,2}$, Ce Feng$^{1,2}$, Yu Zheng$^{1,2}$, Yang Dong$^{1,2}$, Guang-Can Guo$^{1,2}$, Fang-Wen Sun$^{1,2}$

\vspace{5 mm}
\textit{$^1$CAS Key Laboratory of Quantum Information, School of physics, University of Science and Technology of China, Hefei, 230026, People's Republic of China.}\\
\textit{$^2$CAS Center For Excellence in Quantum Information and Quantum Physics, University of Science and Technology of China, Hefei, 230026, People's Republic of China.}

\end{center}


\clearpage
\section{The shape of the antenna}

The whole bowtie structure contains two parts: a small bowtie structure of Au film and a large bowtie structure of copper tap, as shown in Supplementary Figure \ref{figantenna}a. The gap of the small bowtie structure is W$_{gap}$=8 $\mu$m, and the width of the bowtie structure is W$_1$ =160 $\mu$m at the gap. The length of the whole structure $L$ is changed by cutting the copper tape. To optimize the length of the bowtie structure for NV center spin manipulation, we record the Rabi oscillation frequency of the NV center with different bowtie lengths (without the Ag nanowire), as shown in Supplementary Figure \ref{figantenna}b. The maximum Rabi oscillation frequency is obtained with a length of 6.3 cm.

\begin{figure*}[h]
  \centering
  \includegraphics[width=0.85\textwidth]{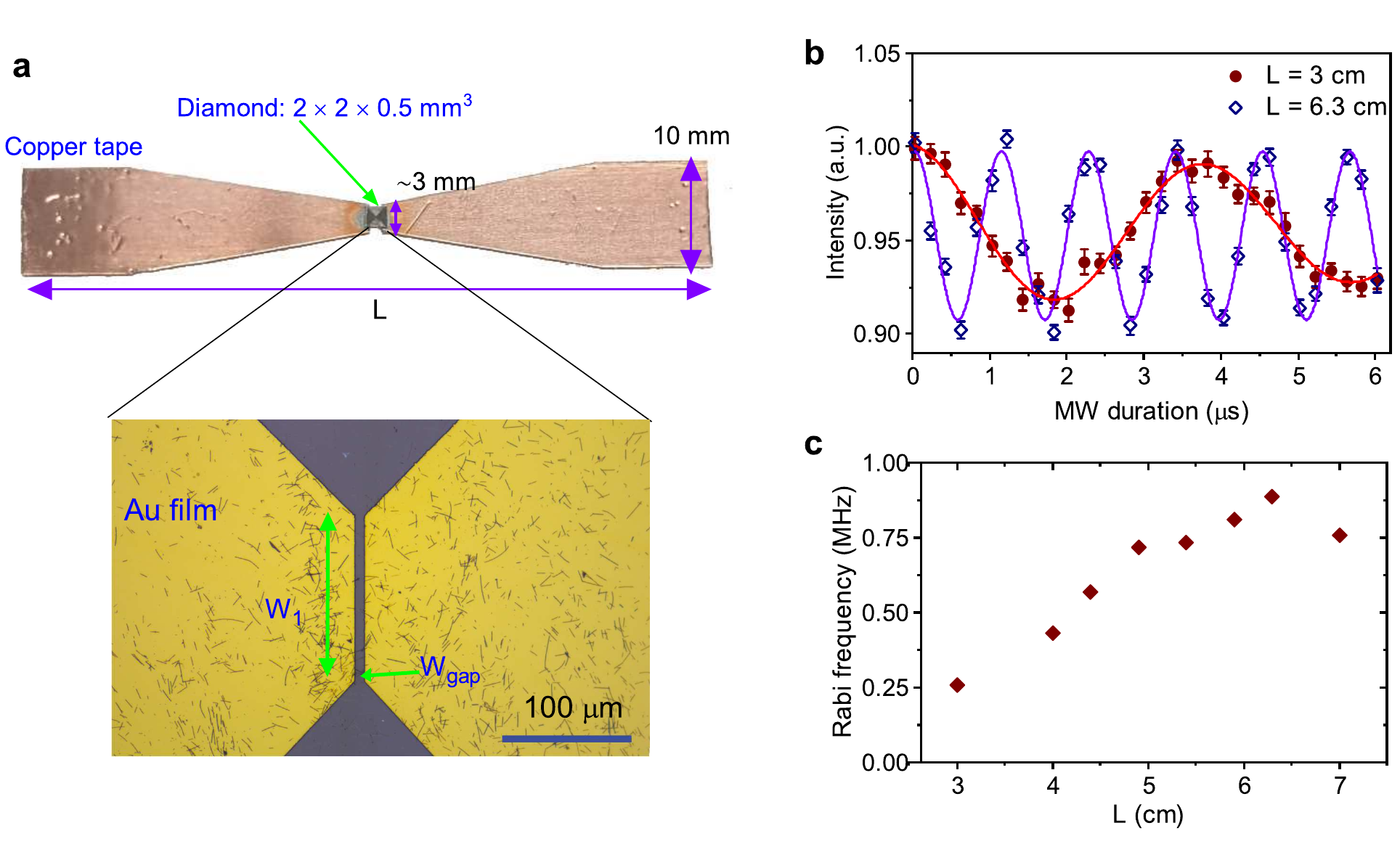}
  \caption{\textbf{Optimizing the nanowire-bowtie structure.} (a) The shape of the nanowire-bowtie antenna. (b) Rabi oscillation of the NV center in the gap of bowtie structure with $L = $3 and 6.3 cm. The power of the free-space microwave is 21 W. (c) The Rabi oscillation frequency changes with $L$.}\label{figantenna}
\end{figure*}

\section{The microwave power dependent ODMR contrast}

The ODMR contrast of NV center is defined as
\begin{equation}\label{eqcontrast}
  C=\frac{I_{m_{s}=0}-I_{m_{s}=\pm 1}}{I_{m_{s}=0}},
\end{equation}
where $I_{m_s=0}$ is the fluorescence intensity with spin state m$_s$ = 0 (no microwave pumping) and $I_{m_{s} =\pm 1}$ is the fluorescence intensity with spin state m$_{s}= \pm 1$ (with microwave pumping). To calibrate the power dependence of the OOMR contrast, we use a ring shape near-field antenna with a high spatial uniformity to pump the spin transition of NV center\cite{MWante-2016rsi}. The ODMR contrast of the NV center is recorded with different microwave amplitude B$_{MW}$, as in Supplementary Figure \ref{figcontrast}. The contrast is then fitted by a saturation function\cite{budker-odmr-2013prb}
\begin{equation}\label{eqsat}
  C = C_{0} \frac{B_{MW}}{B_{MW}+B_{sat}},
\end{equation}
where $C_{0}$ is the saturated contrast and $B_{sat}$ is the saturation microwave amplitude.

 \begin{figure*}[h]
  \centering
  \includegraphics[width=8.5cm]{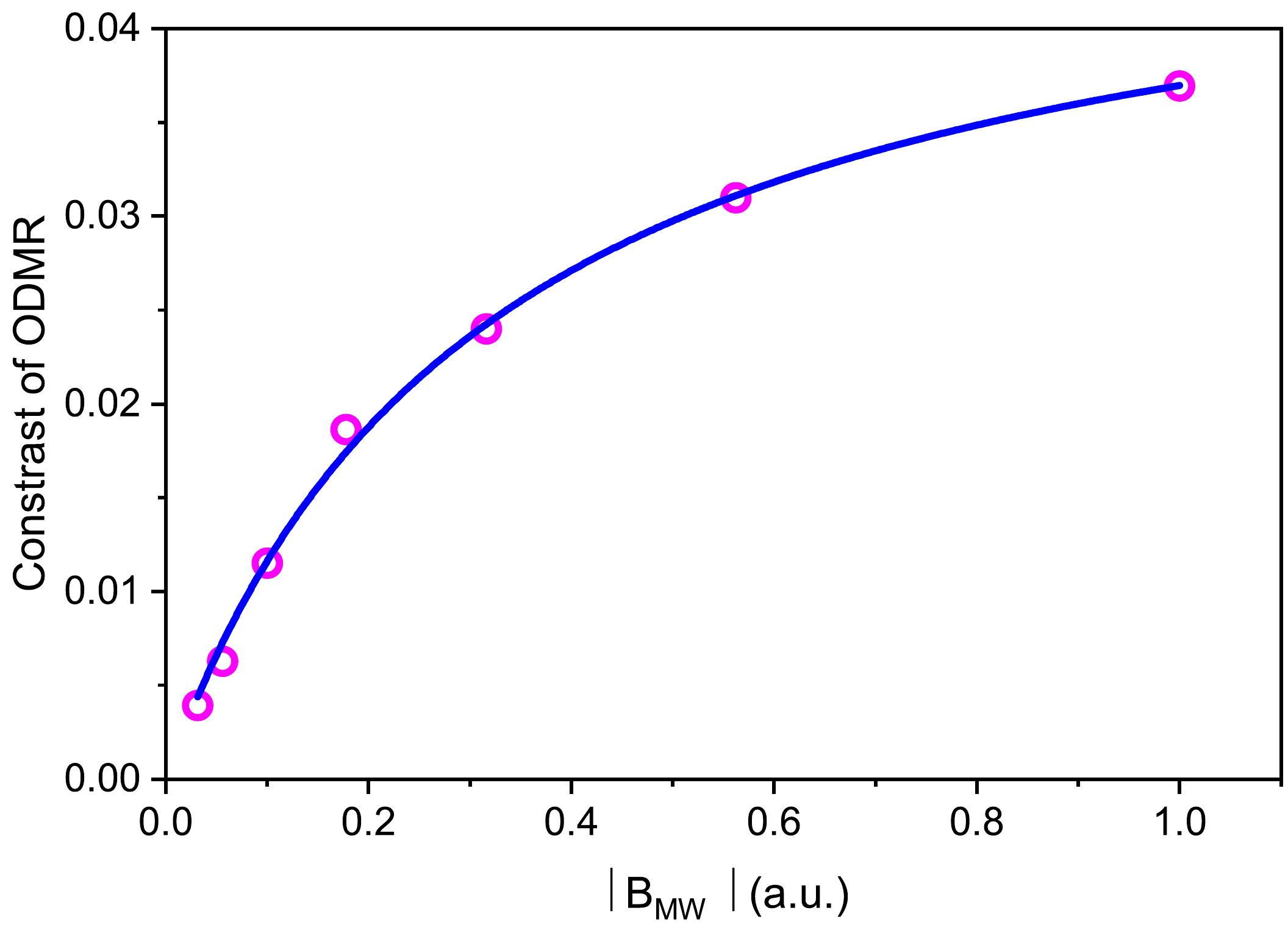}
  \caption{\textbf{The contrast of ODMR signal with different microwave amplitudes.} The solid line is the fit of equation \ref{eqsat}. } \label{figcontrast}
\end{figure*}

\section{The resolution of CSD nanoscopy}

 \begin{figure*}[h]
  \centering
  \includegraphics[width=10cm]{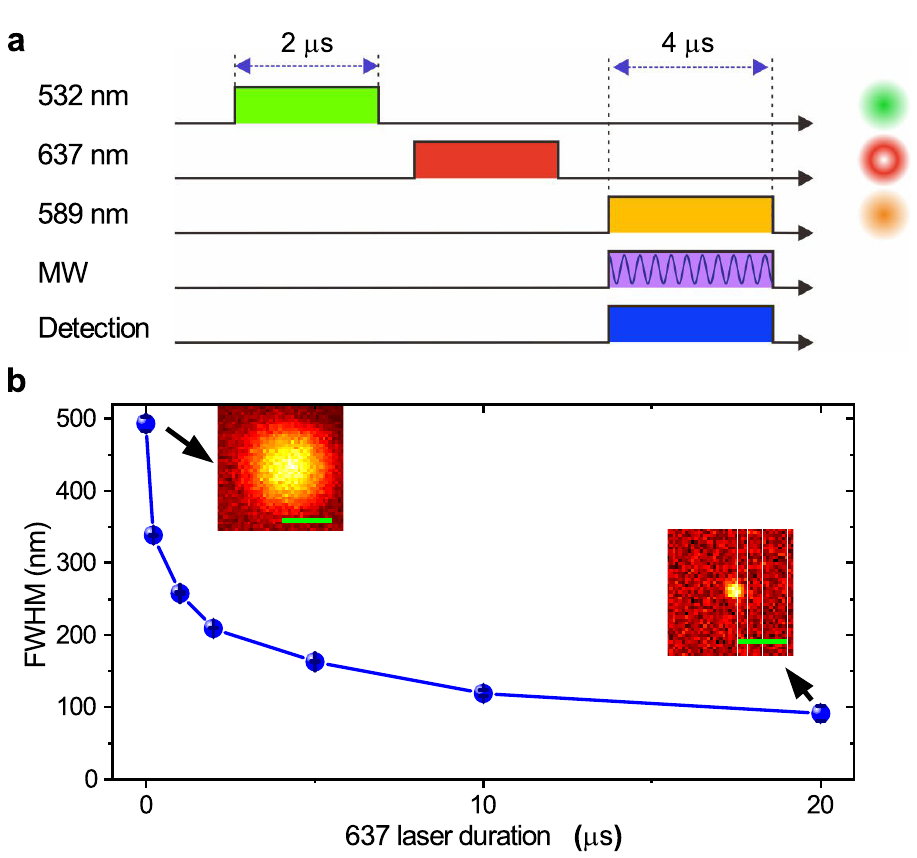}
  \caption{\textbf{The CSD nanoscopy for the diffraction-unlimited ODMR measurement.} (a) The sequence for the CSD nanoscopy. (b) The CSD nanoscopy resolution changes with the 637 nm depletion laser duration. The power of the 637 nm depletion laser is 20 mW here. The inserts show the images of a single NV center. The scale bars in the inserts are 400 nm in length.}\label{figcsd}
\end{figure*}

The CSD nanoscopy is based on the optical manipulation of NV center charge state. Two charge states of NV center is utilized: the negatively charged NV$^{-}$, and the neutrally charged NV$^{0}$.
The conversion between the two charge states can be pumped by laser pulse. And the two charge states can be optically distinguished according to the fluorescence intensity and wavelength. In our experiments, a long pass optical filter is used to block the fluorescence of NV$^{0}$ charge state. Therefore, we treat the NV$^{0}$ as a dark state in the CSD nanoscopy.

The laser and microwave sequence for super-resolution imaging and sensing is shown in Supplementary Figure \ref{figcsd}a. A Gaussian-shaped green laser initializes NV center to the negatively charged NV$^{-}$ state. Then a doughnut-shaped red laser changes the charge state to the neutral NV$^{0}$ charge state. The NV at the center of the doughnut-shaped laser beam remains to be NV$^{-}$. After that, the spin state transition of NV$^{-}$ ground state is pumped by the local microwave field. The charge state and spin state of NV center is detected with a 589 nm laser.

In this work, the power of the 532 and 589 nm lasers are set to 0.2 and 0.05 mW, respectively. The spatial resolution of the CSD nanoscopy increases with the duration and power of the 637 nm doughnut-shaped depletion laser. Here, we set the power of the 637 nm laser to 20 mW. With a duration of 10 $\mu$s, the lateral resolution of NV imaging is improved to approximate 100 nm. It is high enough to map the distribution of the localized microwave field near the Ag nanowire.

\section{The Johnson noise}

The random motion of electron in a conductor will cause electromagnetic fluctuations, known as Johnson noise. It decreases the spin relaxation time of the nearby NV center, and subsequently affects the applications with NV or other solid state spin. We measure the spin relaxation of NV centers at different positions with the nanowire-bowtie antenna. As expected, the Au film significantly decreases the relaxation time T$_{1}$ of NV center. In contrast, the spin relaxation time of NV center under the Ag nanowire does not show significant change. It suggests that the impact of Johnson noise from an Ag nanowire is small. The nanowire-bowtie antenna can be used for spin manipulation without reducing the relaxation time.

\begin{figure*}[h]
  \centering
  \includegraphics[width=8.5cm]{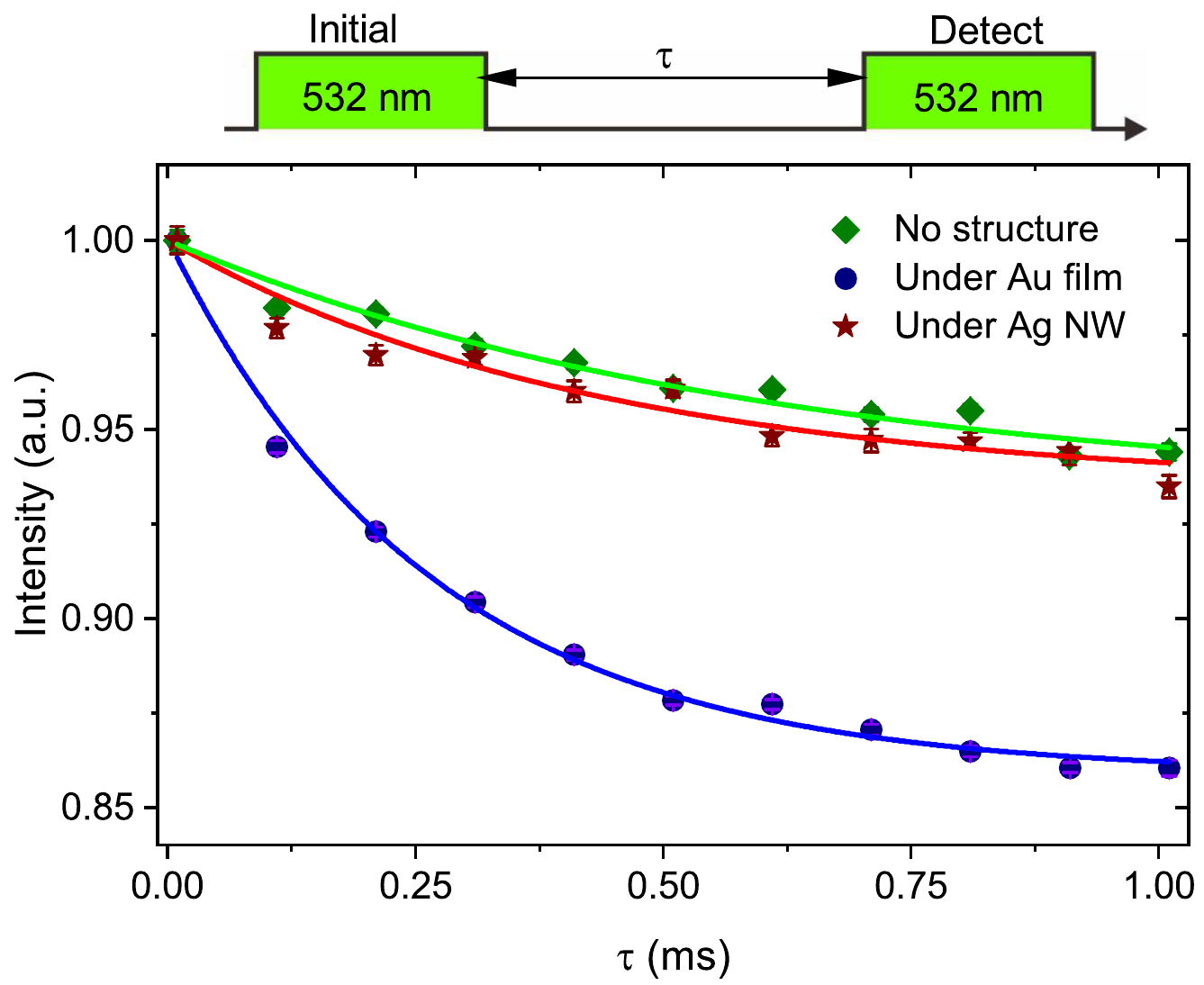}
  \caption{\textbf{The spin relaxation of NV center.} The spin state of NV center is firstly initialized to $m_{s}=0$ with a 532 nm laser pulse. After a decay time $\tau$, the spin state is detected with another 532 nm laser pulse. The fluorescence intensity decreases as the spin state decays to a mixed state.}\label{figt1}
\end{figure*}

\section{The Simulation}

\begin{figure*}[h]
  \centering
  \includegraphics[width=0.8\textwidth]{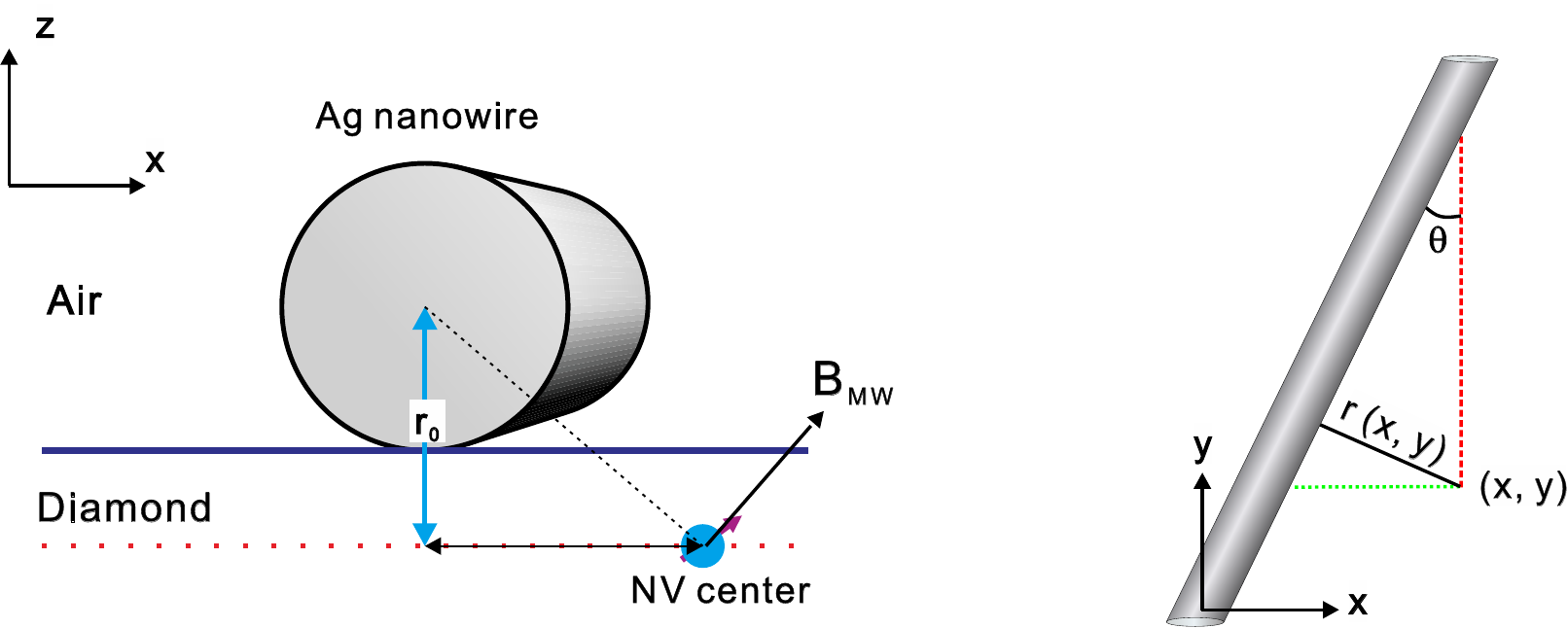}
  \caption{\textbf{The illustration of the microwave that is produced by a straight line current.} }\label{figsimu}
\end{figure*}

To simulate the magnetic component of the microwave field that is produced by a straight line current, we consider an Ag nanowire in the xy plane. The angle between the y axis and the nanowire is $\theta$. The position of nanowire is written as ($x_0$, $y_0$), which follows $y_0 = \cot\theta \cdot x_0$. For a straight line current on the Ag nanowire, the magnetic amplitude of the microwave at position ($x, y$) is
\begin{equation}\label{eqmagam}
  B_{MW}(x,y)\propto \frac{1}{\sqrt{r_0^2+r^2}},
\end{equation}
where $r = \sin\theta \cdot(\cot\theta\cdot x-y)$. $r_0$ is determined by the radius of the nanowire and the depth of NV center. The projections of magnetic vector can be written as:
\begin{align}
B_{MW,x} & =B_{MW}\frac{r_{0}}{\sqrt{r_{0}^2+r^2}}\cdot \cos\theta,\\
B_{MW,y} & =B_{MW}\frac{r_{0}}{\sqrt{r_{0}^2+r^2}}\cdot \sin\theta,\\
B_{MW,z} & =B_{MW}\frac{r}{\sqrt{r_{0}^2+r^2}}.
\end{align}
The amplitude of the microwave component that effectively pumps NVi center spin transition is then written as:
\begin{align}
B_{MW1} & =\sqrt{B_{MW}^2-(-\sqrt{\frac{2}{3}}B_{MW,x}+\sqrt{\frac{1}{3}}B_{MW,z})^2},\\
B_{MW2} & =\sqrt{B_{MW}^2-(\sqrt{\frac{2}{3}}B_{MW,x}+\sqrt{\frac{1}{3}}B_{MW,z})^2},\\
B_{MW3} & =\sqrt{B_{MW}^2-(\sqrt{\frac{2}{3}}B_{MW,y}+\sqrt{\frac{1}{3}}B_{MW,z})^2},\\
B_{MW4} & =\sqrt{B_{MW}^2-(-\sqrt{\frac{2}{3}}B_{MW,y}+\sqrt{\frac{1}{3}}B_{MW,z})^2}.
\end{align}

The fluorescence intensity distribution without microwave is written as $I_0 (x,y)$. The fluorescence of NVi center under microwave pumping is:
\begin{equation}\label{eqfluormw}
  I_{NVi}(x,y)=I_{0}(x,y)\cdot(1-C_{NVi}),
\end{equation}
where $C_{NVi}$ is the ODMR contrast of NVi center. $C_{NVi}$ is determined by the amplitude of $B_{MWi}$, as in equation \ref{eqsat}.

In our experiments, the CSD nanoscopy is applied for the microwave distribution measurement. The point spreading function of the CSD nanoscopy is written as
\begin{equation}\label{eqpsf}
PSF(x,y)=\frac{1}{\sqrt{2\pi \sigma}} e^{-\frac{x^2+y^2}{2\sigma ^2}}.
\end{equation}
The full width at half maximum (FWHM = 2.355 $\sigma$) presents the resolution of the CSD nanoscopy. The detected fluorescence signal is the convolution of the real fluorescence distribution and the point spreading function of the CSD nanoscopy:
\begin{equation}\label{eqfluoconv}
I_{det,NVi}(x,y)=\iint I_{NVi}(x_{1},y_{1})\cdot PSF(x-x_{1},y-y_{1})dx_{1}dy_{1}.
\end{equation}
The convolution of the fluorescence distribution without microwave pumping is also calculated as
\begin{equation}\label{eqfluoconv0}
I_{det,0}(x,y)=\iint I_{0}(x_{1},y_{1})\cdot PSF(x-x_{1},y-y_{1})dx_{1}dy_{1}.
\end{equation}
Then, the detected ODMR signal distribution is simulated as
\begin{equation}\label{eqODMRconv}
C_{det,i}(x,y)=\frac{I_{det,0}(x,y)-I_{det,NVi}(x,y)}{I_{det,0}(x,y)}.
\end{equation}

In the main text, we show that the simulation of a straight line current matches well with the experimental results.


%